\documentclass[twoside,a4paper,11pt]{sca}
\usepackage{graphicx}
\usepackage{hyperref}
\usepackage{movie15}
\usepackage{natbib}  
\newcommand {\nbody}{\textsc{\mbox{nbody6}}}
\newcommand {\nbodypp}{\textsc{\mbox{nbody6\raise.4ex\hbox{\tiny++}}}}
\newcommand {\secnbodypp}{\textsc{\mbox{nbody6\raise.4ex\hbox{\scriptsize++}}}}
\newcommand {\nbodygpu} {\textsc{\mbox{nbody6-gpu}}}
\newcommand {\nbodyppgpu}{\textsc{\mbox{nbody6\raise.4ex\hbox{\tiny++}gpu}}}
\newcommand {\starlab} {\texttt{starlab}}

\newcommand {\Msun} {\mbox{M$_{\odot}$}}

\newcommand {\lmst}     {\mbox{$l_{\mathrm{MST}}$}}
\newcommand {\lref}     {\mbox{$l_{\mathrm{MST}}^{\mathrm{ref}}$}}
\newcommand {\lmass}    {\mbox{$l_{\mathrm{MST}}^{\mathrm{mass}}$}}
\newcommand {\Lmst}     {\mbox{$\Lambda_{\mathrm{MST}}$}}
\newcommand {\gmst}     {\mbox{$\gamma_{\mathrm{MST}}$}}
\newcommand {\gref}     {\mbox{$\gamma_{\mathrm{MST}}^{\mathrm{ref}}$}}
\newcommand {\gmass}    {\mbox{$\gamma_{\mathrm{MST}}^{\mathrm{mass}}$}}
\newcommand {\Gmst}     {\mbox{$\Gamma_{\mathrm{MST}}$}}

\newcommand {\Mmstl}    {\mbox{${\cal{M}}_{\mathrm{MST}}^{\mathrm{\Lambda}}$}}
\newcommand {\Mmstg}    {\mbox{${\cal{M}}_{\mathrm{MST}}^{\mathrm{\Gamma}}$}}

\begin{document}
\pagenumbering{arabic}
\pagestyle{myheadings}
\thispagestyle{empty}
{\flushright\includegraphics[width=\textwidth,bb=90 650 520 700]{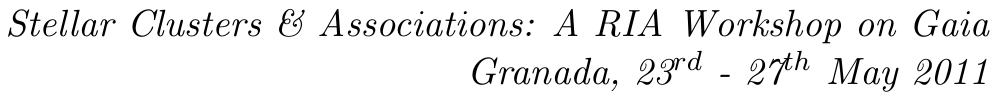}}
\vspace*{0.2cm}
\begin{flushleft}
{\bf {\LARGE
%
Dynamics in Young Star Clusters: From Planets to Massive Stars
%
}\\
\vspace*{1cm}
%
C. Olczak$^{1,2,3}$,
R. Spurzem$^{3,1}$,
Th. Henning$^{2}$,
T. Kaczmarek$^{4}$,
S. Pfalzner$^{4}$,
S. Harfst$^{5}$,
and 
S. Portegies Zwart$^{6}$
%
}\\
\vspace*{0.5cm}
%
$^{1}$
Astronomisches Rechen-Institut (ARI), Zentrum f{\"u}r Astronomie Universit{\"a}t Heidelberg, M{\"o}nchhofstrasse 12-14, 69120 Heidelberg, Germany\\
$^{2}$
Max-Planck-Institut f{\"u}r Astronomie (MPIA), K{\"o}nigstuhl 17, 69117 Heidelberg, Germany\\
$^{3}$
National Astronomical Observatories of China, Chinese Academy of Sciences (NAOC/CAS), 20A Datun Lu, Chaoyang District, Beijing 100012, China\\
$^{4}$
Max-Planck-Institut f{\"u}r Radioastronomie, Auf dem H{\"u}gel 7, 53121 Bonn, Germany\\
$^{5}$
Technische Universit{\"a}t Berlin, Zentrum f{\"u}r Astronomie und Astrophysik, Hardenbergstra{\ss}e 36, 10623 Berlin, Germany\\
$^{6}$ Sterrewacht Leiden, Leiden University, Postbus 9513, 2300 RA Leiden, The Netherlands
%
\end{flushleft}
%
\markboth{
Dynamics in Young Star Clusters: From Planets to Massive Stars
}{ 
%
C. Olczak et al.
%
}
\thispagestyle{empty}
\vspace*{0.4cm}
\begin{minipage}[l]{0.09\textwidth}
\ 
\end{minipage}
\begin{minipage}[r]{0.9\textwidth}
\vspace{1cm}
\section*{Abstract}{\small
%
The young star clusters we observe today are the building blocks of a new generation of stars and planets in our Galaxy and beyond. Despite
their fundamental role we still lack knowledge about the conditions under which star clusters form and the impact of these often harsh
environments on the evolution of their stellar and substellar members.

We demonstrate the vital role numerical simulations play to uncover both key issues. Using dynamical models of different star cluster environments we
show the variety of effects stellar interactions potentially have. Moreover, our significantly improved measure of mass segregation
reveals that it can occur rapidly even for star clusters without substructure. This finding is a critical step to resolve the controversial debate on
mass segregation in young star clusters and provides strong constraints on their initial conditions.
%
\normalsize}
\end{minipage}
%
%
%
\section{Introduction \label{intro}}
According to current knowledge, planetary systems form from the accretion discs around young stars. These young stars are in most cases not isolated,
but are part of a cluster \citep[e.g.][]{2003ARA&A..41...57L,2009ApJS..181..321E}. Densities in these cluster environments vary considerably, spanning
a range of 10\,pc$^{-3}$ (e.g. $\eta$ Chameleontis) to $10^6$\,pc$^{-3}$ (e.g. Arches Cluster). Though it is known that discs disperse on a time-scale
of 1-10\,Myr \citep{2001ApJ...553L.153H,2002astro.ph.10520H,2006ApJ...638..897S,2008ApJ...672..558C} and that in dense clusters
($n$\,$\ge$\,$10^3$\,pc$^{-3}$) the disc frequency seems to be lower in the core \citep[e.g.][]{2007ApJ...660.1532B}, it is an open question as to how
far interactions with the surrounding stars influence the planet formation in clusters of different densities. An encounter between a circumstellar
disc and a nearby passing star can lead to a significant loss of mass and angular momentum from the disc. While such isolated encounters have been
studied in a large variety
\citep{1993ApJ...408..337H,1993MNRAS.261..190C,1994ApJ...424..292O,1995ApJ...455..252H,1996MNRAS.278..303H,1997MNRAS.287..148H,2004ApJ...602..356P,2005ApJ...629..526P,2006ApJ...653..437M,2007ApJ...656..275M,2008A&A...487..671K},
only a few numerical studies have directly investigated the effect of stellar encounters on circumstellar discs in a dense cluster environment
\citep{2001MNRAS.325..449S,2006ApJ...641..504A}.

The dynamical evolution of a star cluster leaves a variety of imprints in the phase space of its stellar population which are good tracers of the
\emph{dynamical} age of the cluster. One of the most widely discussed aspects is that of mass segregation. Due to energy equipartition -- hence via
two-body encounters -- the more massive particles tend to settle towards the cluster centre over time while the lower-mass particles are
preferentially pushed to the outer parts \citep{1969ApJ...158L.139S,1983ApJ...271...11F,1995MNRAS.272..772S,2007MNRAS.374..703K}. However, it is a
much more challenging task to identify mass segregation observationally in real objects than theoretically from `clean' numerical simulations. The
investigation of mass segregation in young stellar systems is of particular interest for a deeper understanding of the star formation process.

\section{Star-Disc Encounters in Young Star Clusters}

\subsection{Method}

We follow the idea of \cite{2001MNRAS.325..449S}, combining a simulation of the dynamics of a cluster to determine the interaction parameters of close
encounters between stars in the cluster with results from studies of isolated star-disc encounter simulations. Throughout this work we assume that
initially all stars are surrounded by protoplanetary discs. This is justified by observations that reveal disc fractions of nearly 100\,\% in very
young star clusters \citep[e.g.][]{2000AJ....120.1396H,2000AJ....120.3162L,2001ApJ...553L.153H,2005astro.ph.11083H}.

The dynamical cluster models contain only single stellar components without considering embedded gas. The simulations were performed with $\nbodypp$,
$\nbodygpu$ \citep{1999JCoAM.109..407S,2003grav.book.....A,2008LNP...760....1A}, and $\starlab$
\citep[][]{1996ASPC...90..413M,2001MNRAS.321..199P,2003IAUS..208..331H}. The details of the numerics can be found in \cite[][2011, in
prep]{2006ApJ...642.1140O}.

The encounter-induced transport of mass and angular momentum in protoplanetary discs is calculated via fit formulae based on simulations of star-disc
encounters \citep{2006ApJ...642.1140O,2006A&A...454..811P,2007A&A...462..193P}. These quantify upper limits due to the restriction to low-mass discs
and co-planar, prograde, and parabolic orbits, which are the most perturbing. A simplified prescription terms stars that have lost more than 90\,\% of
their initial disc mass as ``discless''. The results presented here apply to a scenario where all stars have initially a disc size of 150\,AU. A
lerger range of models has been investigated in the cited publications.

\subsection{Results}

\begin{figure}
  \centering
  \includegraphics[height=0.49\textwidth,angle=-90]{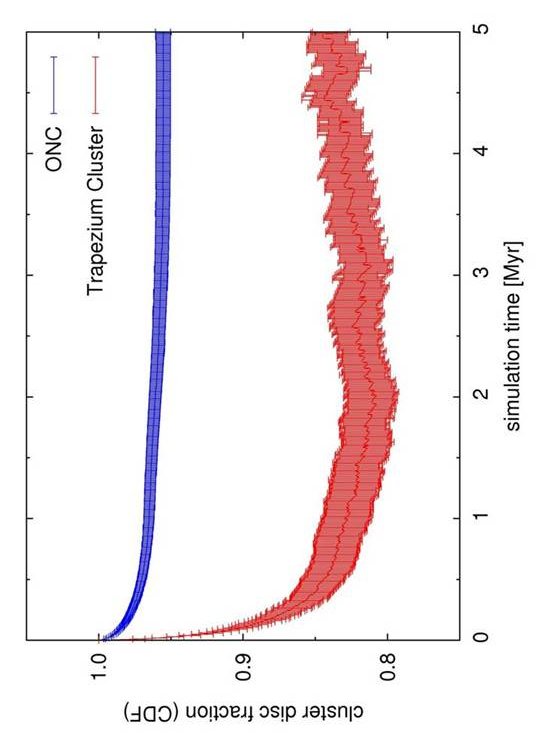}
  \includegraphics[height=0.49\textwidth,angle=-90]{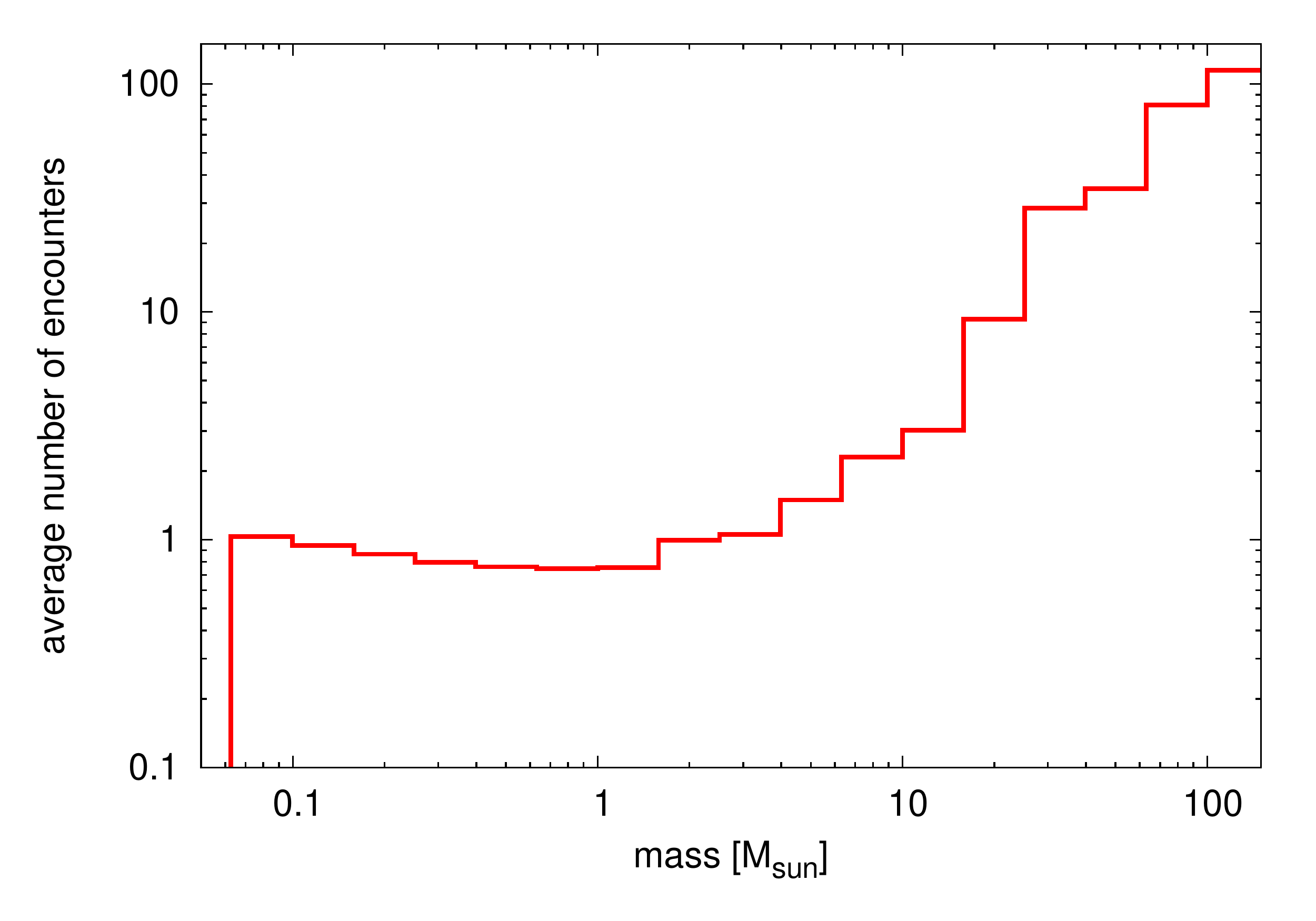}
  \caption{\emph{Left:} Fraction of star-disc system over time in the entire ONC ($R = 2.5$\,pc) and its dense core, the Trapezium cluster ($R =
    0.3$\,pc). \emph{Right:} Average cumulative number of encounters vs. stellar mass after 1\,Myr of dynamical evolution.}
  \label{fig:onc__disc_mass_loss}
\end{figure}
From a numerical model of the Orion Nebula Cluster (ONC) we demonstrate that the encounter-induced disc-mass loss becomes significant in its dense
core, known as the ``Trapezium Cluster'' (Fig.~\ref{fig:onc__disc_mass_loss}, left). Up to 20\,\% of the discs can be destroyed by gravitational
interactions within a radius of 0.3\,pc (compared to $\sim$5\,\% in the entire cluster of 2.5\,pc radius).

The massive stars in the centre of such a stellar cluster act as gravitational foci for the lower mass stars \citep{2006A&A...454..811P}. This becomes
evident from the average number of encounters as a function of stellar mass (Fig.~\ref{fig:onc__disc_mass_loss}, right): the number of disc-perturbing
interactions is nearly constant for low- and intermediate-mass stars but increases largely for high-mass stars. Because discs are most affected when
the masses of the interacting stars are unequal \citep{2006ApJ...642.1140O,2007ApJ...656..275M} massive stars dominate the encounter-induced disc-mass
loss in star clusters like the ONC. For low-mass stars the mass-loss occurs through few strong encounter events, whereas the disc of high-mass stars
is removed via a steady nibbling by many encounters with stars of lower mass.

\begin{figure}
  \includegraphics[height=0.5\textwidth,angle=-90]{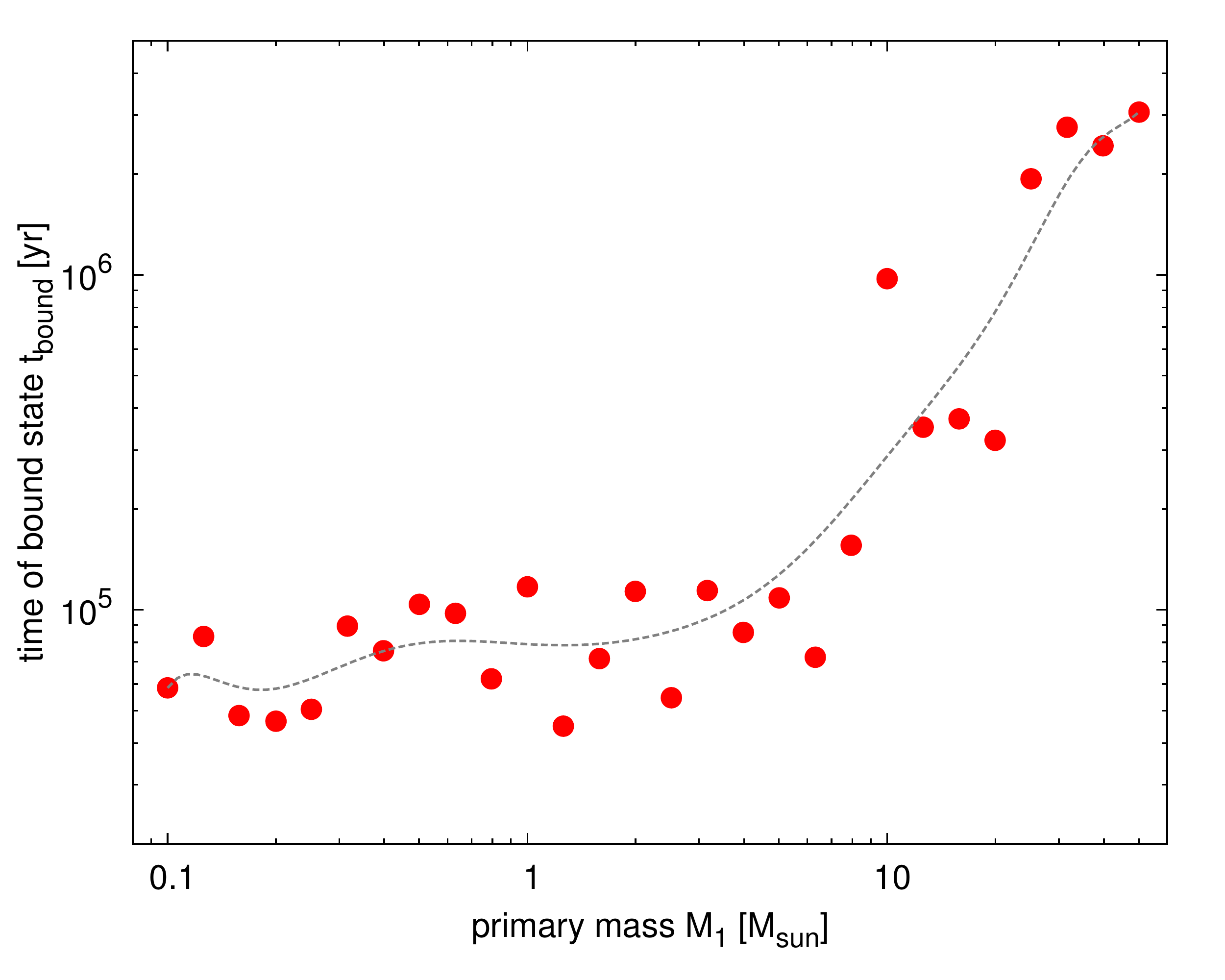}
  \includegraphics[height=0.5\textwidth,angle=-90]{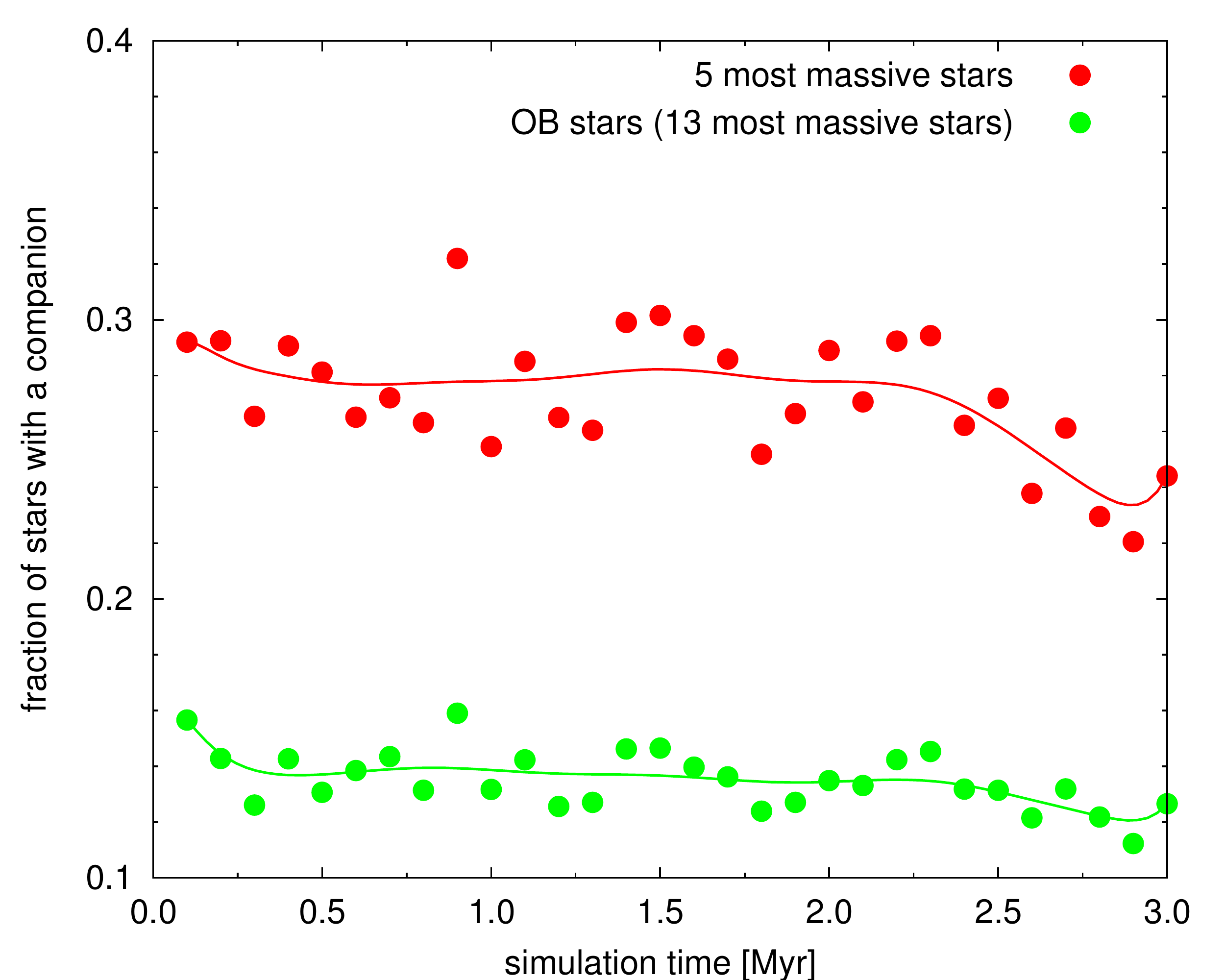}
  \caption{\emph{Left:} Average duration of a bound state, $t_{bound}$, as a function of the primary mass $M_1$. \emph{Right:} Fraction of the 5 (red)
    and 13 (green) most massive stars that are part of an encounter-induced binary as a function of cluster age. The lines are Bezier curves of the
    data.}
  \label{fig:onc__tbs}
\end{figure}
The dynamically outstanding role of high-mass stars does also affect the stellar multiplicity in the ONC. It turns out that the most massive star
($M_1^* = 50\,\Msun$) has on average of the order of 200 so-called ``capturing encounters'' -- encounter events with an eccentricity $\epsilon <$ 1 --
in the first 5 Myr while a star with $M_1^* = 1\,\Msun$ has less than one. These interactions lead to the formation of relatively stable
configurations termed as ``transient bound systems'' (TBS). The left plot in Fig.~\ref{fig:onc__tbs} shows a striking increase of $t_{bound}$ towards
more massive stars by more than one order of magnitude. These TBS translate directly into an observed apparent binarity (Fig.~\ref{fig:onc__tbs},
right): of the five most massive stars in the ONC on average 30\,\% would appear the be in a binary system at the cluster's current age of
$\sim$1\,Myr. The average for all 13 OB stars in the ONC would be more than 10\,\%.


\begin{figure}
  \centering
  \includegraphics[height=0.49\textwidth,angle=-90]{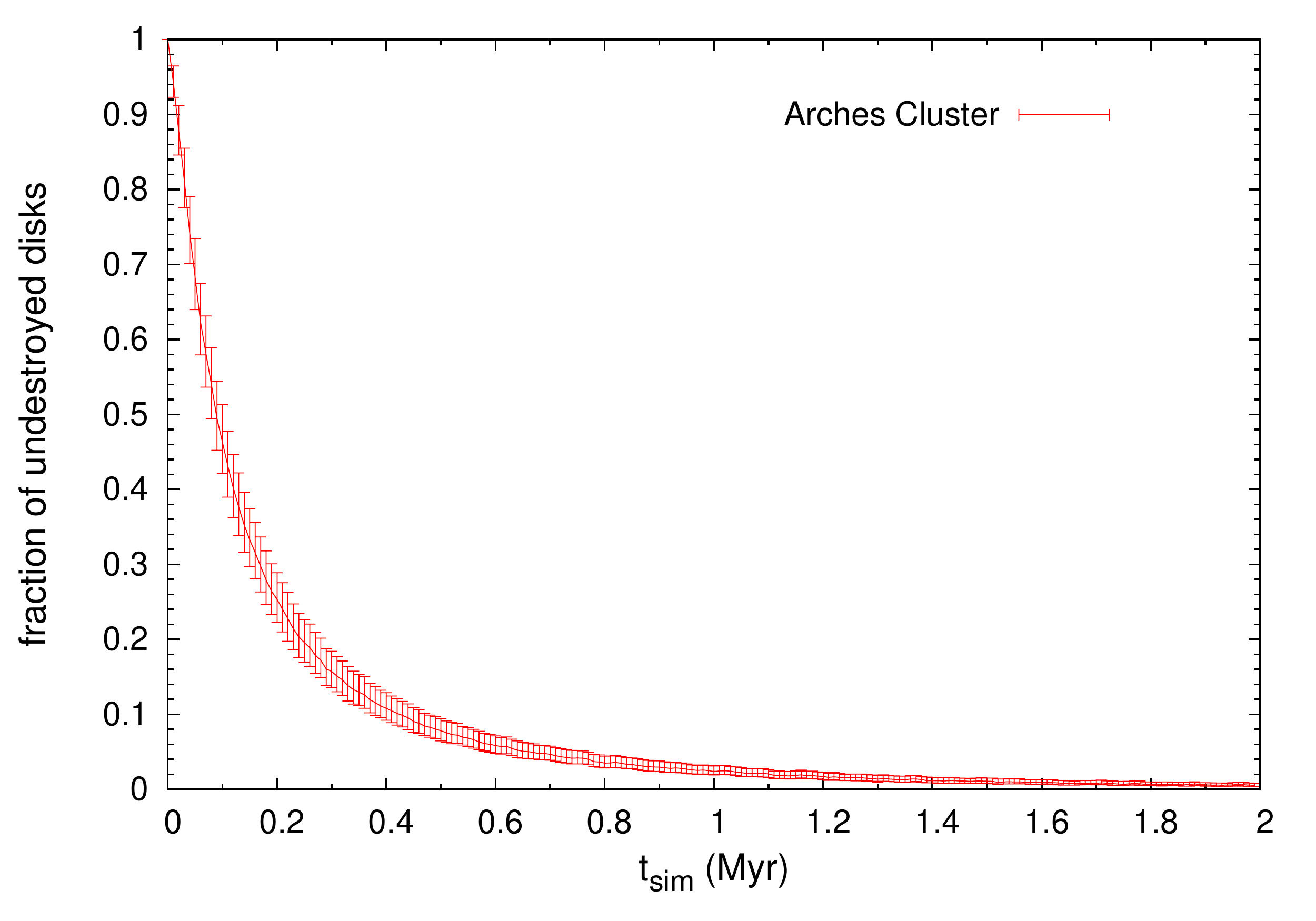}
  \includegraphics[height=0.49\textwidth,angle=-90]{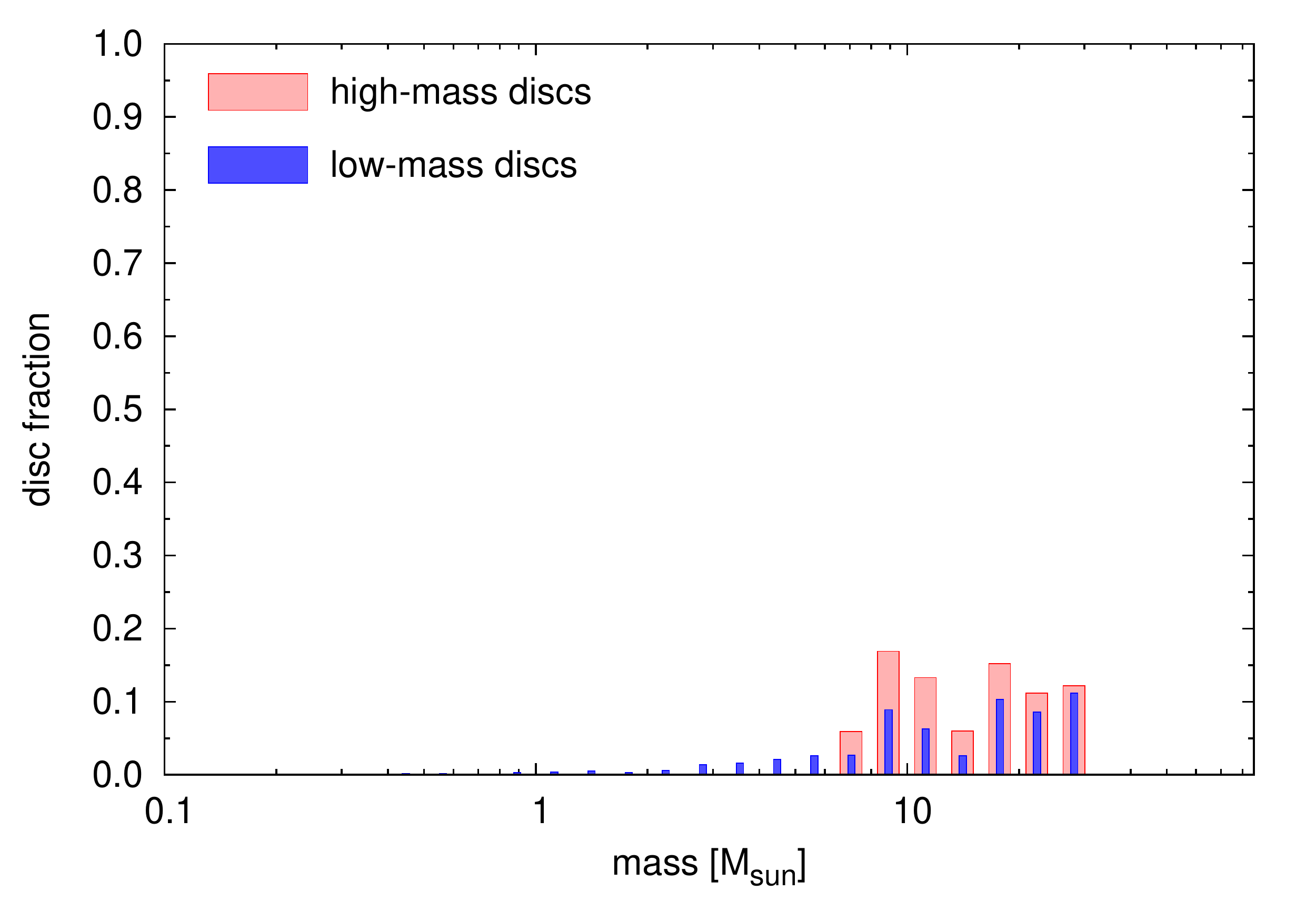}
  \caption{\emph{Left:} Fraction of star-disc system over time in the entire Arches cluster. \emph{Right}: Average disc fraction vs. stellar mass
    after 2\,Myr of dynamical evolution.}
  \label{fig:arches__disc_mass_loss}
\end{figure}  

The encounter-induced disc-mass loss becomes much more apparent in more extreme environments. Using a numerical model of the Arches cluster, one of
the densest and most massive clusters in our Milky Way \citep{2002ApJ...581..258F,2005ApJ...628L.113S}, we find that at its current age of
$\sim$2.5\,Myr \citep{2004ApJ...611L.105N,2008A&A...478..219M} nearly all circumstellar discs could have been destroyed just due to gravitational
interactions (Fig.~\ref{fig:arches__disc_mass_loss}, left). A striking result is that the low fraction of members that could retain their discs to
some degree at an age of 2\,Myr is populated by stars with masses between 2\,$\Msun$ and 30\,$\Msun$ (Fig.~\ref{fig:arches__disc_mass_loss}, right).

\section{Mass Segregation in Young Star Clusters}

\subsection{Method}

As a proxy for mass segregation we extend the method $\Mmstl$ developed by \citet{2009MNRAS.395.1449A} \citep[see
also][]{2004MNRAS.348..589C,2006A&A...449..151S}. In summary, the authors use the minimum spanning tree (MST), the graph which connects all vertices
within a given sample with the lowest possible sum of edges and no closed loops \citep{1969JRSS...18...54G}. The length of the MST, $\lmst$, is a
measure of the concentration or compactness of a given sample of vertices. Mass segregation of a stellar system of size $N$ is quantified by comparing
$\lmst$ of the $n$ most massive stars, $\lmass$, with the average $\lmst$ of $k$ sets of $n$ random cluster stars, $<\lref>$, and its standard
deviation, $\Delta\lref$.

Our method $\Mmstg$ involves a crucial modification of $\Mmstl$ that boosts its sensitivity: we do not use directly the \emph{sum} of the edges
$\lmst$ as a measure yet their \emph{geometric mean} $\gmst$,
\begin{equation}
  \gmst = \bigg( \prod_{i=1}^n e_i \bigg)^{1/n} = \exp{ \left[ \frac{1}{n} \sum_{i=1}^n \ln{e_i} \right] } \,,
\end{equation}
and its associated geometric standard deviation $\Delta\gmst$,
where $e_i$ are the $n$ MST edges. We obtain the new measure $\Gmst$ via a proper normalisation:
\begin{equation}
  \Gmst       = \frac{ \gref }{ \gmass } \,, \quad
  \Delta\Gmst = \Delta\gref \,.
  \label{eq:Gmst}
\end{equation}
The geometrical mean acts as an intermediate-pass that damps contributions from extreme edge lengths very effectively, hence significantly reduces the
contribution of any ``outlier''.

\subsection{Results}


\begin{figure}
  \centering
  \includegraphics[height=0.49\linewidth,angle=-90]{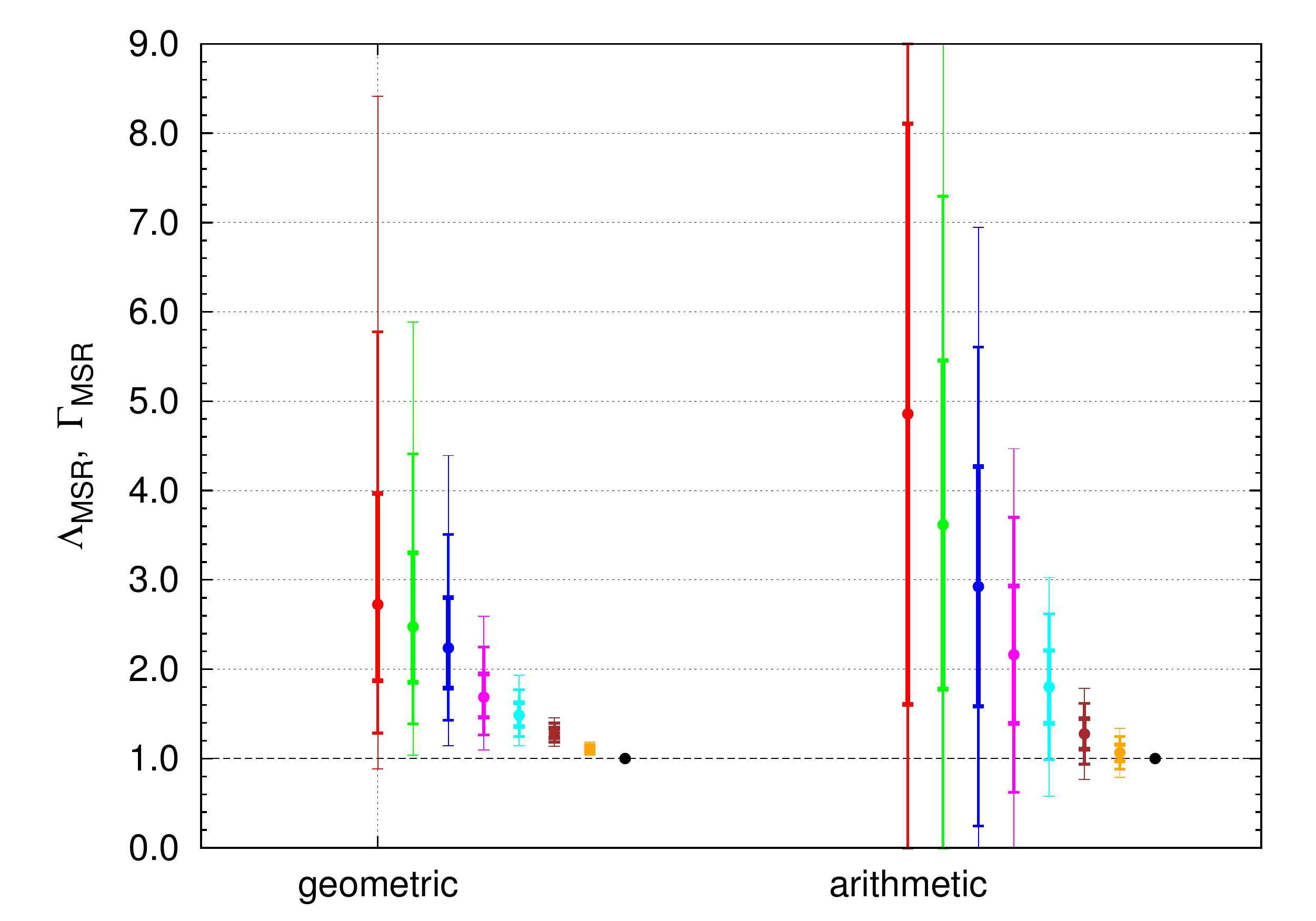}
  \includegraphics[height=0.50\linewidth,angle=-90]{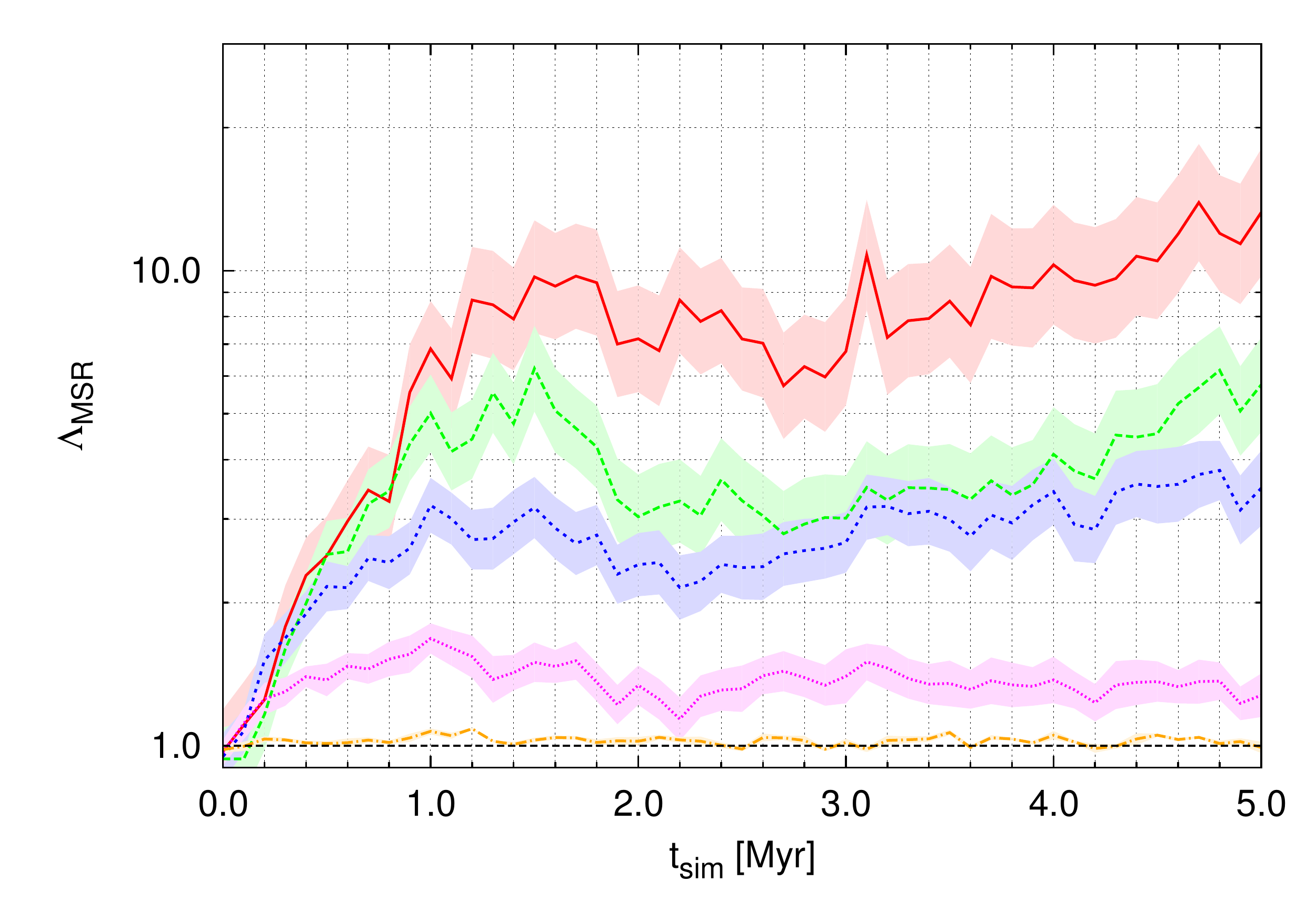}
  \caption{\emph{Left:} $\Gmst$ and $\Lmst$ for the 5, 10, 20, 50, 100, 200, 500, 1000 most massive stars of an initially mass segregated cluster of
    1k stars and $S = 0.3$. The error bars and line thickness mark the $1\sigma$, $2\sigma$, and $3\sigma$ uncertainties. \emph{Right:} $\Gmst$ of the
    5, 10, 20, 50, and 500 most massive stars over time for a cluster with 1k members and cold initial conditions ($Q=0.1$). The filled regions
    indicate 1$\sigma$ uncertainties.}
  \label{fig:mass_segregation}
\end{figure}

Using the method of \citet{2008MNRAS.385.1673S} we have created initially mass-segregated star clusters to verify our algorithm. In the left plot of
Fig.~\ref{fig:mass_segregation} we compare our improved measure $\Gmst$ (``geometric'') with $\Lmst$ (``arithmetic''). We find that our measure of
mass segregation, $\Gmst$, detects an intermediate degree of mass segregation, $S = 0.3$, with at least $3\sigma$ significance while $\Lmst$ provides
only a very weak $1\sigma$ significance.

We have used this highly improved sensitivity to follow the dynamical mass segregation of a numerical star cluster model. The initial configuration is
based on our ONC model as described in previous publications \citep[e.g.][]{2010A&A...509A..63O}. Note that we use a spherically symmetric model with
a smooth density distribution without any substructure. However, our stellar system is initially collapsing, starting from a virial ratio $Q =
0.1$. The simulations were carried out until a physical age of 5\,Myr, corresponding to 13.5 N-body time units. On the right-hand side of
Fig.~\ref{fig:mass_segregation} we see that $\Gmst$ increases rapidly within the first 2\,Myr (or 5.5 N-body time units) and then saturates. The
different mass groups are clearly separated, inversely correlated with the sample size.

\section{Discussion and Conclusions}

From numerical simulations of star cluster dynamics and star-disc encounters we find that pure gravitational interactions of disc-surrounded stars can
lead to a significant depletion of their circumstellar matter. The disc-mass loss increases with cluster density but remains unaffected by the size of
the stellar population. In moderately dense clusters like the ONC it is the massive stars that dominate the encounter-induced disc-mass loss by
gravitational focusing of low-mass stars.

Massive stars have a much higher probability to be involved in a capturing encounter in such a cluster environment than solar-mass stars. In the ONC
at least 10\,\% of the OB stars form a (transient) bound state due to dynamical capture that would be observed as a binary. The properties of the
these systems strongly depend on the cluster age with a development towards smaller periastra, higher mass ratios and longer duration of the bound
states.

In extreme environments like the Arches cluster the pairwise interaction of all stars leads nearly to a complete destruction of the entire disc
population. The preferential survivability of discs around B-type stars in the Arches cluster implies that these could be the best candidates for
detecting planetary systems in Galactic starburst clusters.

We have developed a significantly improved method to measure mass segregation based on the minimum spanning tree (MST). In general, using only the ten
to twenty most massive stars $\Gmst$ provides a robust and sensitive measure of mass segregation for the entire population of star clusters in our
Galaxy.  In particular, very low degrees of mass segregation can be detected in massive clusters like NGC~3603 that consist of 10k or more stars.

We note that mass segregation in a collapsing, intermediate-size stellar cluster of 1k stars can occur very quickly, i.e. within only a few crossing
times. This finding demonstrates that rapid mass segregation (in terms of dynamical time scale) does \emph{not} require substructure.
%
%
\small  
%
\section*{Acknowledgments}   
%
CO and RS acknowledge support by NAOC CAS through the Silk Road Project, and by Global Networks and Mobility Program of the University of Heidelberg
(ZUK 49/1 TP14.8 Spurzem). CO appreciates funding by the German Research Foundation (DFG), grant OL~350/1-1. RS is funded by the Chinese Academy of
Sciences Visiting Professorship for Senior International Scientists, Grant Number 2009S1-5. We have partly used the special supercomputers at the
Center of Information and Computing at National Astronomical Observatories, Chinese Academy of Sciences, funded by Ministry of Finance of People’s
Republic of China under the grant ZDY Z2008−2.

We thank S. Aarseth for providing the highly sophisticated N-body code $\nbody$ (and its GPU extension) and greatly appreciate his support.

%
%
%
%
%

\bibliographystyle{aa}
\bibliography{mnemonic,Olczak_C_ref}

\begin{thebibliography}{46}
\expandafter\ifx\csname natexlab\endcsname\relax\def\natexlab#1{#1}\fi

\bibitem[{{Aarseth}(2003)}]{2003grav.book.....A}
{Aarseth}, S. 2003, {Gravitational N-body Simulations} (Cambridge, Cambridge
  University Press, 2003, 430 p.)

\bibitem[{{Aarseth}(2008)}]{2008LNP...760....1A}
{Aarseth}, S.~J. 2008, in Lecture Notes in Physics, Berlin Springer Verlag,
  Vol. 760, Lecture Notes in Physics, Berlin Springer Verlag, ed. S.~J.
  {Aarseth}, C.~A. {Tout}, \& R.~A. {Mardling}, 1--4020

\bibitem[{{Adams} {et~al.}(2006){Adams}, {Proszkow}, {Fatuzzo}, \&
  {Myers}}]{2006ApJ...641..504A}
{Adams}, F.~C., {Proszkow}, E.~M., {Fatuzzo}, M., \& {Myers}, P.~C. 2006, apj,
  641, 504

\bibitem[{{Allison} {et~al.}(2009){Allison}, {Goodwin}, {Parker}, {Portegies
  Zwart}, {de Grijs}, \& {Kouwenhoven}}]{2009MNRAS.395.1449A}
{Allison}, R.~J., {Goodwin}, S.~P., {Parker}, R.~J., {et~al.} 2009, mnras, 395,
  1449

\bibitem[{{Balog} {et~al.}(2007){Balog}, {Muzerolle}, {Rieke}, {Su}, {Young},
  \& {Megeath}}]{2007ApJ...660.1532B}
{Balog}, Z., {Muzerolle}, J., {Rieke}, G.~H., {et~al.} 2007, apj, 660, 1532

\bibitem[{{Cartwright} \& {Whitworth}(2004)}]{2004MNRAS.348..589C}
{Cartwright}, A. \& {Whitworth}, A.~P. 2004, mnras, 348, 589

\bibitem[{{Clarke} \& {Pringle}(1993)}]{1993MNRAS.261..190C}
{Clarke}, C.~J. \& {Pringle}, J.~E. 1993, mnras, 261, 190

\bibitem[{{Currie} {et~al.}(2008){Currie}, {Kenyon}, {Balog}, {Rieke}, {Bragg},
  \& {Bromley}}]{2008ApJ...672..558C}
{Currie}, T., {Kenyon}, S.~J., {Balog}, Z., {et~al.} 2008, apj, 672, 558

\bibitem[{{Evans} {et~al.}(2009){Evans}, {Dunham}, {J{\o}rgensen}, {Enoch},
  {Mer{\'{\i}}n}, {van Dishoeck}, {Alcal{\'a}}, {Myers}, {Stapelfeldt},
  {Huard}, {Allen}, {Harvey}, {van Kempen}, {Blake}, {Koerner}, {Mundy},
  {Padgett}, \& {Sargent}}]{2009ApJS..181..321E}
{Evans}, N.~J., {Dunham}, M.~M., {J{\o}rgensen}, J.~K., {et~al.} 2009, apjs,
  181, 321

\bibitem[{{Farouki} {et~al.}(1983){Farouki}, {Hoffman}, \&
  {Salpeter}}]{1983ApJ...271...11F}
{Farouki}, R.~T., {Hoffman}, G.~L., \& {Salpeter}, E.~E. 1983, apj, 271, 11

\bibitem[{{Figer} {et~al.}(2002){Figer}, {Najarro}, {Gilmore}, {Morris}, {Kim},
  {Serabyn}, {McLean}, {Gilbert}, {Graham}, {Larkin}, {Levenson}, \&
  {Teplitz}}]{2002ApJ...581..258F}
{Figer}, D.~F., {Najarro}, F., {Gilmore}, D., {et~al.} 2002, apj, 581, 258

\bibitem[{Gower \& Ross(1969)}]{1969JRSS...18...54G}
Gower, J.~C. \& Ross, G. J.~S. 1969, Journal of the Royal Statistical Society.
  Series C (Applied Statistics), 18, pp. 54

\bibitem[{{Haisch} {et~al.}(2000){Haisch}, {Lada}, \&
  {Lada}}]{2000AJ....120.1396H}
{Haisch}, K.~E., {Lada}, E.~A., \& {Lada}, C.~J. 2000, aj, 120, 1396

\bibitem[{{Haisch} {et~al.}(2001){Haisch}, {Lada}, \&
  {Lada}}]{2001ApJ...553L.153H}
{Haisch}, Jr., K.~E., {Lada}, E.~A., \& {Lada}, C.~J. 2001, apjl, 553, L153

\bibitem[{{Hall}(1997)}]{1997MNRAS.287..148H}
{Hall}, S.~M. 1997, mnras, 287, 148

\bibitem[{{Hall} {et~al.}(1996){Hall}, {Clarke}, \&
  {Pringle}}]{1996MNRAS.278..303H}
{Hall}, S.~M., {Clarke}, C.~J., \& {Pringle}, J.~E. 1996, mnras, 278, 303

\bibitem[{{Heller}(1993)}]{1993ApJ...408..337H}
{Heller}, C.~H. 1993, apj, 408, 337

\bibitem[{{Heller}(1995)}]{1995ApJ...455..252H}
{Heller}, C.~H. 1995, apj, 455, 252

\bibitem[{{Hillenbrand}(2002)}]{2002astro.ph.10520H}
{Hillenbrand}, L.~A. 2002, ArXiv Astrophysics e-prints

\bibitem[{{Hillenbrand}(2005)}]{2005astro.ph.11083H}
{Hillenbrand}, L.~A. 2005, ArXiv Astrophysics e-prints

\bibitem[{{Hut}(2003)}]{2003IAUS..208..331H}
{Hut}, P. 2003, in IAU Symposium, Vol. 208, Astrophysical Supercomputing using
  Particle Simulations, ed. {J.~Makino \& P.~Hut}, 331--+

\bibitem[{{Khalisi} {et~al.}(2007){Khalisi}, {Amaro-Seoane}, \&
  {Spurzem}}]{2007MNRAS.374..703K}
{Khalisi}, E., {Amaro-Seoane}, P., \& {Spurzem}, R. 2007, mnras, 374, 703

\bibitem[{{Kley} {et~al.}(2008){Kley}, {Papaloizou}, \&
  {Ogilvie}}]{2008A&A...487..671K}
{Kley}, W., {Papaloizou}, J.~C.~B., \& {Ogilvie}, G.~I. 2008, aap, 487, 671

\bibitem[{{Lada} \& {Lada}(2003)}]{2003ARA&A..41...57L}
{Lada}, C.~J. \& {Lada}, E.~A. 2003, araa, 41, 57

\bibitem[{{Lada} {et~al.}(2000){Lada}, {Muench}, {Haisch}, {Lada}, {Alves},
  {Tollestrup}, \& {Willner}}]{2000AJ....120.3162L}
{Lada}, C.~J., {Muench}, A.~A., {Haisch}, Jr., K.~E., {et~al.} 2000, aj, 120,
  3162

\bibitem[{{Martins} {et~al.}(2008){Martins}, {Hillier}, {Paumard},
  {Eisenhauer}, {Ott}, \& {Genzel}}]{2008A&A...478..219M}
{Martins}, F., {Hillier}, D.~J., {Paumard}, T., {et~al.} 2008, aap, 478, 219

\bibitem[{{McMillan}(1996)}]{1996ASPC...90..413M}
{McMillan}, S.~L.~W. 1996, in Astronomical Society of the Pacific Conference
  Series, Vol.~90, The Origins, Evolution, and Destinies of Binary Stars in
  Clusters, ed. {E.~F.~Milone \& J.-C.~Mermilliod}, 413--+

\bibitem[{{Moeckel} \& {Bally}(2006)}]{2006ApJ...653..437M}
{Moeckel}, N. \& {Bally}, J. 2006, apj, 653, 437

\bibitem[{{Moeckel} \& {Bally}(2007)}]{2007ApJ...656..275M}
{Moeckel}, N. \& {Bally}, J. 2007, apj, 656, 275

\bibitem[{{Najarro} {et~al.}(2004){Najarro}, {Figer}, {Hillier}, \&
  {Kudritzki}}]{2004ApJ...611L.105N}
{Najarro}, F., {Figer}, D.~F., {Hillier}, D.~J., \& {Kudritzki}, R.~P. 2004,
  apjl, 611, L105

\bibitem[{{Olczak} {et~al.}(2010){Olczak}, {Pfalzner}, \&
  {Eckart}}]{2010A&A...509A..63O}
{Olczak}, C., {Pfalzner}, S., \& {Eckart}, A. 2010, aap, 509, A260000+

\bibitem[{{Olczak} {et~al.}(2006){Olczak}, {Pfalzner}, \&
  {Spurzem}}]{2006ApJ...642.1140O}
{Olczak}, C., {Pfalzner}, S., \& {Spurzem}, R. 2006, apj, 642, 1140

\bibitem[{{Ostriker}(1994)}]{1994ApJ...424..292O}
{Ostriker}, E.~C. 1994, apj, 424, 292

\bibitem[{{Pfalzner}(2004)}]{2004ApJ...602..356P}
{Pfalzner}, S. 2004, apj, 602, 356

\bibitem[{{Pfalzner} \& {Olczak}(2007)}]{2007A&A...462..193P}
{Pfalzner}, S. \& {Olczak}, C. 2007, aap, 462, 193

\bibitem[{{Pfalzner} {et~al.}(2006){Pfalzner}, {Olczak}, \&
  {Eckart}}]{2006A&A...454..811P}
{Pfalzner}, S., {Olczak}, C., \& {Eckart}, A. 2006, aap, 454, 811

\bibitem[{{Pfalzner} {et~al.}(2005){Pfalzner}, {Umbreit}, \&
  {Henning}}]{2005ApJ...629..526P}
{Pfalzner}, S., {Umbreit}, S., \& {Henning}, T. 2005, apj, 629, 526

\bibitem[{{Portegies Zwart} {et~al.}(2001){Portegies Zwart}, {McMillan}, {Hut},
  \& {Makino}}]{2001MNRAS.321..199P}
{Portegies Zwart}, S.~F., {McMillan}, S.~L.~W., {Hut}, P., \& {Makino}, J.
  2001, mnras, 321, 199

\bibitem[{{Scally} \& {Clarke}(2001)}]{2001MNRAS.325..449S}
{Scally}, A. \& {Clarke}, C. 2001, mnras, 325, 449

\bibitem[{{Schmeja} \& {Klessen}(2006)}]{2006A&A...449..151S}
{Schmeja}, S. \& {Klessen}, R.~S. 2006, aap, 449, 151

\bibitem[{{Sicilia-Aguilar} {et~al.}(2006){Sicilia-Aguilar}, {Hartmann},
  {Calvet}, {Megeath}, {Muzerolle}, {Allen}, {D'Alessio}, {Mer{\'{\i}}n},
  {Stauffer}, {Young}, \& {Lada}}]{2006ApJ...638..897S}
{Sicilia-Aguilar}, A., {Hartmann}, L., {Calvet}, N., {et~al.} 2006, apj, 638,
  897

\bibitem[{{Spitzer}(1969)}]{1969ApJ...158L.139S}
{Spitzer}, L.~J. 1969, apjl, 158, L139+

\bibitem[{{Spurzem}(1999)}]{1999JCoAM.109..407S}
{Spurzem}, R. 1999, Journal of Computational and Applied Mathematics, 109, 407

\bibitem[{{Spurzem} \& {Takahashi}(1995)}]{1995MNRAS.272..772S}
{Spurzem}, R. \& {Takahashi}, K. 1995, mnras, 272, 772

\bibitem[{{Stolte} {et~al.}(2005){Stolte}, {Brandner}, {Grebel}, {Lenzen}, \&
  {Lagrange}}]{2005ApJ...628L.113S}
{Stolte}, A., {Brandner}, W., {Grebel}, E.~K., {Lenzen}, R., \& {Lagrange},
  A.-M. 2005, apjl, 628, L113

\bibitem[{{{\v S}ubr} {et~al.}(2008){{\v S}ubr}, {Kroupa}, \&
  {Baumgardt}}]{2008MNRAS.385.1673S}
{{\v S}ubr}, L., {Kroupa}, P., \& {Baumgardt}, H. 2008, mnras, 385, 1673

\end{thebibliography}

\end{document}